\documentstyle[epsf]{mn}

\newif\ifAMStwofonts

\def\kms{\mbox{\,km s$^{-1}$}}

\def\Ms{\mbox{${\rm \,M_{\odot}}$}}

\def\mdot{\mbox{\,${\rm \dot{M}}$}}
\def\msunyr{\mbox{\,${\rm M_{\odot}\, yr^{-1}}$}}

\def\ccc{\mbox{\,cm$^{-3}$}}

\ifoldfss
  \ifCUPmtlplainloaded \else
    \NewTextAlphabet{textbfit} {cmbxti10} {}
    \NewTextAlphabet{textbfss} {cmssbx10} {}
    \NewMathAlphabet{mathbfit} {cmbxti10} {} 
    \NewMathAlphabet{mathbfss} {cmssbx10} {} 
  \fi
  \ifAMStwofonts
    \ifCUPmtlplainloaded \else
      \NewSymbolFont{upmath} {eurm10}
      \NewSymbolFont{AMSa} {msam10}
      \NewMathSymbol{\upi}     {0}{upmath}{19}
      \NewMathSymbol{\umu}     {0}{upmath}{16}
      \NewMathSymbol{\upartial}{0}{upmath}{40}
      \NewMathSymbol{\leqslant}{3}{AMSa}{36}
      \NewMathSymbol{\geqslant}{3}{AMSa}{3E}

      \let\leq=\leqslant 
       
    \fi
  \fi
\fi 

\ifnfssone
  \newmathalphabet{\mathit}
  \addtoversion{normal}{\mathit}{cmr}{m}{it}
  \addtoversion{bold}{\mathit}{cmr}{bx}{it}
  \newmathalphabet{\mathbfit} 
  \addtoversion{normal}{\mathbfit}{cmr}{bx}{it}
  \addtoversion{bold}{\mathbfit}{cmr}{bx}{it}
  \newmathalphabet{\mathbfss} 
  \addtoversion{normal}{\mathbfss}{cmss}{bx}{n}
  \addtoversion{bold}{\mathbfss}{cmss}{bx}{n}
  \ifAMStwofonts
    \ifCUPmtlplainloaded \else
      \UseAMStwoboldmath
      \makeatletter
      \new@mathgroup\upmath@group
      \define@mathgroup\mv@normal\upmath@group{eur}{m}{n}
      \define@mathgroup\mv@bold\upmath@group{eur}{b}{n}
      \edef\UPM{\hexnumber\upmath@group}
      \new@mathgroup\amsa@group
      \define@mathgroup\mv@normal\amsa@group{msa}{m}{n}
      \define@mathgroup\mv@bold\amsa@group{msa}{m}{n}
      \edef\AMSa{\hexnumber\amsa@group}
      \makeatother
      \mathchardef\upi="0\UPM19
      \mathchardef\umu="0\UPM16
      \mathchardef\upartial="0\UPM40
      \mathchardef\leqslant="3\AMSa36
      \mathchardef\geqslant="3\AMSa3E

      \let\leq=\leqslant 

    \fi
  \fi
\fi 

\ifnfsstwo
  \DeclareMathAlphabet{\mathbfit}{OT1}{cmr}{bx}{it}
  \SetMathAlphabet\mathbfit{bold}{OT1}{cmr}{bx}{it}
  \DeclareMathAlphabet{\mathbfss}{OT1}{cmss}{bx}{n}
  \SetMathAlphabet\mathbfss{bold}{OT1}{cmss}{bx}{n}
  \ifAMStwofonts
    \ifCUPmtlplainloaded \else
      \DeclareSymbolFont{UPM}{U}{eur}{m}{n}
      \SetSymbolFont{UPM}{bold}{U}{eur}{b}{n}
      \DeclareSymbolFont{AMSa}{U}{msa}{m}{n}
      \DeclareMathSymbol{\upi}{0}{UPM}{"19}
      \DeclareMathSymbol{\umu}{0}{UPM}{"16}
      \DeclareMathSymbol{\upartial}{0}{UPM}{"40}
      \DeclareMathSymbol{\leqslant}{3}{AMSa}{"36}
      \DeclareMathSymbol{\geqslant}{3}{AMSa}{"3E}

      \let\leq=\leqslant 

    \fi
  \fi
\fi 

\ifCUPmtlplainloaded \else
  \ifAMStwofonts \else 
    \def\upi{\pi}
    \def\umu{\mu}
    \def\upartial{\partial}
  \fi
\fi

\title{A distorted radio shell in the young supernova SN\,1986J}
\author[M.A.\ P\'erez-Torres et al.]
  {M.A.~P\'erez-Torres,$^1$ 
   A.~Alberdi,$^2$
   J.M.~Marcaide,$^3$
   J.C.~Guirado,$^3$
   L.~Lara,$^2$
   \newauthor   
   F.~Mantovani,$^1$
   E.~Ros,$^4$
   K.~Weiler$^5$ \\
$^1$Istituto di Radioastronomia -- CNR,
Via P. Gobetti 101, I-40129 Bologna, Italy \\
$^2$Instituto de Astrof\'{\i}sica de Andaluc\'{\i}a, CSIC, Apdo.
Correos 3004, E-18080 Granada, Spain \\
$^3$Departamento de Astronom\'{\i}a y Astrof\'{\i}sica, 
Universidad de Valencia, E-46100 Burjassot, Spain \\
$^4$Max-Planck-Institut f\"ur Radioastronomie, Auf dem H\"ugel 69,
D-53121 Bonn, Germany \\
$^5$Remote Sensing Division, Naval Research Laboratory, Code 7213,
Washington, DC 20375-5320, US  \\
}

\date{Accepted 
      Received 
      in original form 
     }

\pagerange{\pageref{firstpage}--\pageref{lastpage}}
\pubyear{2002}

\begin{document}

\maketitle

\label{firstpage}

\begin{abstract}
We report here on 5 GHz global very-long-baseline interferometry (VLBI) observations
of SN\,1986J, 16 yr after its explosion. 
We obtained a high-resolution image of the 
supernova, which shows a distorted shell of radio emission, indicative  
of a deformation of the shock front.
The angular size of the shell is $\sim4.7\, {\rm mas}$, corresponding
to a linear size of $\sim6.8 \times 10^{17}\, {\rm cm}$ for a distance of 
9.6 Mpc to NGC~891. 
The average speed of the shell has decreased from $\sim$7400\,\kms\, in 1988.74
down to about $6300\, {\rm km\,s^{-1}}$ in 1999.14, 
indicative of a mild deceleration in the expansion of SN\,1986J. 
Assuming a standard density profile for the progenitor
wind ($\rho_{\rm cs}\, \propto\, r^{-s}, s=2$), 
the swept-up mass by the shock front is $\sim$2.2\Ms. 
This large swept-up mass, coupled with the mild deceleration 
suffered by the supernova, suggests that the
mass of the hydrogen-rich envelope ejected at explosion  
was $\ga12$\,\Ms.  Thus, the supernova 
progenitor must have kept intact most of its hydrogen-rich envelope 
by the time of explosion, which favours a single, massive star progenitor scenario. 
We find a flux density for SN\,1986J of $\sim$7.2 mJy at the 
observing frequency of 5 GHz, 
which results in a radio luminosity of $\sim1.4\,\times\,10^{37}\,{\rm erg\, s^{-1}}$
for the frequency range $10^7$--$10^{10}$ Hz  
($\alpha =-0.69; S_\nu\, \propto\, \nu^{\alpha})$.
We detect four bright knots that delineate the shell structure, and
an absolute minimum of emission, which we tentatively identify with
the centre of the supernova explosion. 
If this is the case, SN\,1986J has then suffered an asymmetric expansion.
We suggest that this asymmetry is due to the collision 
of the supernova ejecta with an anisotropic, clumpy (or filamentary) medium. 
\end{abstract}

\begin{keywords}
Techniques: interferometric -- supernovae: individual: SN\,1986J --
ISM: supernova remnants - Radio continuum: stars -- Galaxies: individual: NGC\,891
\end{keywords}

\section{Introduction}
\label{sec:intro}

Type II supernovae (SNe) are associated with massive stars that 
have expelled  slow, dense winds during their supergiant phase.
The stellar explosion drives a shock into this wind, 
at speeds as high as 20000 ${\rm km\,s^{-1}}$ and temperatures of $\sim\,10^9$\,K.  
In addition, a reverse shock propagates back into the stellar envelope 
at speeds of 500-1000 ${\rm km\,s^{-1}}$ relative to the expanding ejecta.  
This is the so-called standard interaction model
(hereafter SIM; Chevalier 1982, Nadyozhin 1985), and
radio, optical, and X-ray emission from Type II supernovae 
have been usually interpreted within this model.
The outgoing shock forms a high-energy-density shell 
that is responsible for the production of
synchrotron radio emission, while the reverse shock
accounts for the optical and soft X-ray emission. 

SN\,1986J in NGC\,891 is one of the most radio luminous SNe ever discovered. 
Indeed, at a distance of $\approx9.6$ Mpc (Tully 1998),
it had a peak luminosity at 5 GHz about 8 and 13 times greater 
than SN\,1979C and SN\,1993J, respectively. 
The precise date of its explosion is not known, but 
on the basis of the available radio and optical data 
SN\,1986J was estimated to have exploded around the end of 1982, 
or the beginning of 1983 
(Rupen et al. 1987, Chevalier 1987, Weiler, Panagia \& Sramek 1990).  
Based upon its large radio luminosity, 
Weiler, Panagia, \& Sramek \cite{weiler90} suggested that
the progenitor star was probably a red giant with
a main-sequence mass of 20 -- 30\,\Ms\, that had lost material
rapidly (\mdot$\,\ga2\,\times\,10^{-4}\,$\msunyr) in a dense stellar
wind. 
VLBI observations by Bartel et al. \cite{bartel91} showed that the radio
structure of SN1986J had the form of a shell, in agreement with 
expectations from the SIM model, with 
the minimum of emission located approximately at its center. 

SN\,1986J is peculiar in several respects.
(i) Bartel et al. \cite{bartel91}  showed the presence of protrusions 
in the brightness distribution of the supernova. 
This finding was interpreted as evidence of  deviation 
from spherical symmetry. 
Another supernova for which considerable asymmetry in its 
radio structure has been found is 
SN1987A (Gaensler et al. 1997, and references
therein).
Those cases contrast with the remarkable 
spherically symmetric expansion
observed for the radio shell of SN\,1993J 
(Marcaide et al. 1995, 1997; Bartel et al. 2000); 
(ii) the radio light curves for SN\,1986J were not well fitted
within the SIM for radio supernovae, which led Weiler et al. (1990)  to
invoke the existence of a mixed medium
(thermal absorbers and non-thermal emitters), or significant 
filamentation in the circumstellar medium; 
(iii) optical spectra obtained in 1986 (Rupen et al. 1987) and 
in 1989 (Leibundgut et al. 1991) showed optical narrow lines with 
FWHM $\approx$ 500 - 700 ${\rm km\,s^{-1}}$. 
In the SIM, the optical and X-ray emission arise in the reverse shock, 
which  moves with a speed close to that of the forward shock, i.e., 
several thousands ${\rm km\,s^{-1}}$.
Since the optical velocities are far lower than those determined 
from VLBI (radio) observations at similar epochs, these narrow lines
question the validity of the SIM
to successfully explain the observed optical and X-ray emission from SN\,1986J.
Chugai \cite{chugai93} proposed a different model which overcomes
the problems of the SIM and seems plausible for SN\,1986J.
In his model, the supernova envelope is not colliding 
with a smooth distribution of wind material, but with a clumpy 
one, and the bulk of the observed X-rays originates in the
shocked dense wind clumps. Therefore, the narrow-line
($\Delta v\,\sim\,500\,{\rm km\,s^{-1}}$) material
is not directly related to the shock wave (which
moves at much higher speeds), but to the 
speed of the shock-excited dense clouds.
More recently, Chugai \& Belous \cite{chugai99} have shown
that the evolution of the radio light curves of SN1986J 
can be well interpreted by free-free absorption in a clumpy wind. 
The possible presence of clumps of radio absorbing plasma is also consistent
with the proposal by Weiler et al. (1990) for a filamentary
structure in the circumstellar medium.  

Unfortunately, VLBI observations of SN\,1986J have been so scarce since
its explosion that only one high-resolution radio image for SN\,1986J exists
(Bartel et al. 1991). 
In this paper, we use very sensitive 5 GHz global VLBI observations 
taken on 21 February 1999 to obtain the second high-resolution 
radio image of SN\,1986J, 16 yr after its explosion. 
We present a brief report of the VLBI observations
in section \ref{sec:obs}, present and discuss our results in 
section \ref{sec:results}, and summarize
our conclusions in section \ref{sec:conclusions}.

\section[]{VLBI Observations and Image Processing}
\label{sec:obs}

We used archival global VLBI data for SN\,1986J.
The supernova was observed at a frequency of 5 GHz from 17:00 UT on 21 February 1999 to
04:50 UT on 22 February 1999, using a very sensitive VLBI array that
included the following antennas (diameter, location):
The whole VLBA (25m, 10 identical antennas across the US), 
phased-VLA (130m-equivalent, NM, US), Effelsberg (100m, Germany), 
Medicina and Noto (32m each, Italy), and Onsala (20m, Sweden). 
The telescopes received both
left- and right-hand circular polarizations (LCP and RCP)  
which, after correlation, were combined to obtain the total
intensity mage presented in this paper.
Effelsberg had technical problems (damaged gear) and thus could not 
take part in the observations.
Onsala only recorded in LCP mode, and therefore its data
were not used to obtain our total intensity VLBI image. 

The observations were made with a bandwidth of 64 MHz. 
The data were correlated at the VLBA Correlator of the National
Radio Astronomy Observatory (NRAO) in Socorro (NM, US). 
The correlator used a pre-averaging time of 3 s. 
Since SN\,1986J was expected to be very faint, at a level of a few mJy, 
the observations were carried out in phase-reference mode. 
SN\,1986J and the nearby ICRF source, 3C\,66A, 
were alternately observed during the 12-hr experiment. 
The observations consisted of $\sim120$\, s scans
on SN1986J and of $\sim70\,$ s scans on 3C\,66A, 
plus a few additional seconds of antenna slew time to make 
a duty cycle of 190 s.
3C\,66A was also observed as amplitude calibrator for SN\,1986J. 
The source 3C\,84 was observed as a fringe finder. 

\begin{figure*}
\mbox{\epsfxsize=16.0cm \epsffile{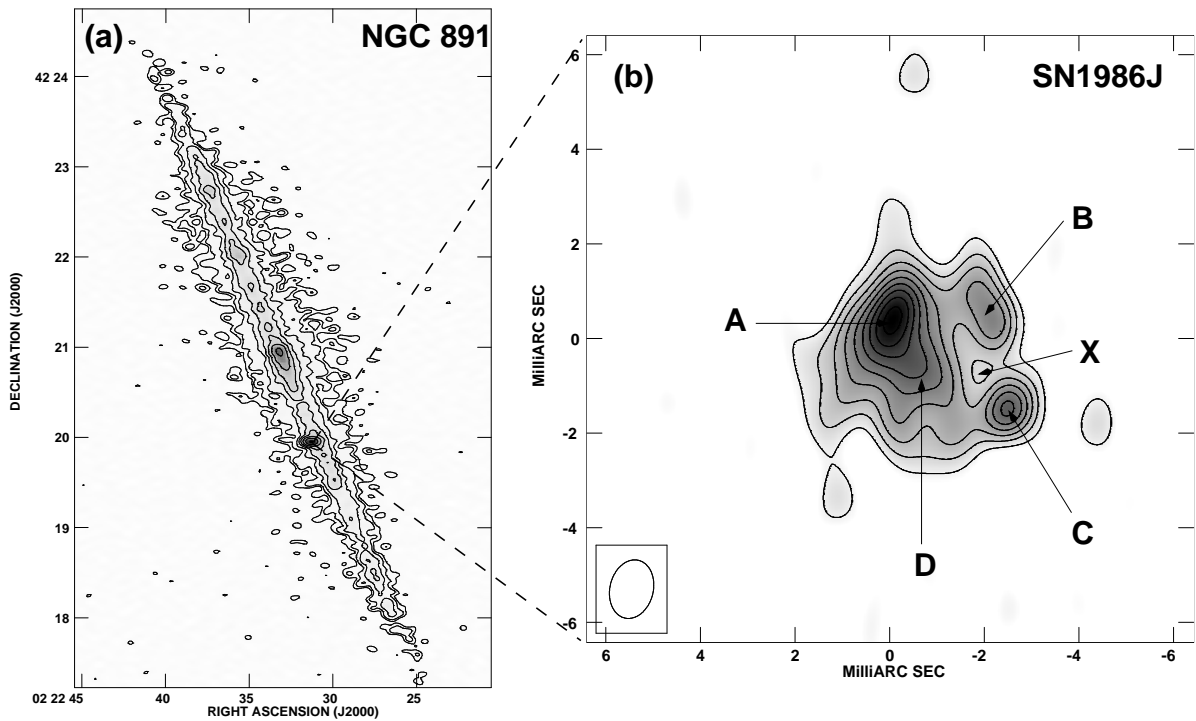}}
\caption{ 
	{\bf (a)} 
	Hybrid map of the galaxy NGC~891 and its supernova SN\,1986J 
	made with the Very Large Array (VLA) at 
	the same frequency (5 GHz) and epoch of the global VLBI observations
        on February 21, 1999. 
	The contours are (3,5,10,20,40,80,160,300) $\times$ 25 $\mu$Jy beam$^{-1}$, 
	the root-mean-square (rms) noise off-source. 
	The peak of brightness of the map corresponds to the supernova 
	and is $\sim$ 8.2 mJy beam$^{-1}$. 
	The dimensions (full width at half maximum) of the restoring beam 
	are $7.4\times4.0$ arcsec$^2$, with the beam's major axis 
	oriented along a position angle  of 88$\fdg$ 
	{\bf (b)}
	Global Very-Long-Baseline Interferometry (VLBI) hybrid map of SN\,1986J  
        on February 21, 1999. 
	The contours are (3,5,7,9,11,13,15,17,19) $\times$ 56 $\mu$Jy beam$^{-1}$, 
	the rms noise off-source. 
	The peak of brightness of the map is 1.13 mJy beam$^{-1}$ and 
	the restoring beam (bottom left in the image) is $1.3 \times 0.9$ mas$^2$ 
	at a position angle of -13$\fdg$4. 
	In both panels, north is up and east is left.
}	 
\label{sn86j}
\end{figure*}

The correlator data were analyzed using the Astronomical Image
Processing System (AIPS).
The visibility amplitudes were calibrated using the system temperature
and gain information provided for each telescope.
The instrumental phase and delay offsets among the 8-MHz baseband
converters in each antenna were corrected using a phase calibration 
determined from observations of 3C\,84.
The data for the calibrator 3C\,66A were then fringe-fitted in a standard manner.
We exported the 3C\,66A data into the Caltech imaging program DIFMAP
(Shepherd et al. 1995) for mapping purposes. 
3C\,66A showed a flux of $\sim0.95$\,Jy at 5 GHz, 
and displayed a one-sided core-jet structure at mas scales, with the jet extending
southwards up to $\sim28$\,mas.
The final source model obtained for 3C\,66A was then included as an 
input model in a new fringe-fitting search for 3C\,66A. 
In this way, the solutions obtained were structure-free.
The phases, delays, and delay-rates determined for  
3C\,66A were then interpolated and applied to the source SN\,1986J. 
The SN\,1986J data were then transferred into the DIFMAP program.
Standard self-calibration techniques and purely uniform weighting 
were used to achieve maximum resolution in the hybrid maps shown 
in Fig. \ref{sn86j}.  

We note two powerful aspects of the use of the phase-reference technique:
(i) it effectively increases the integration time on a source from minutes to 
hours, thus allowing the detection and imaging of very faint objects
(e.g. Beasley and Conway 1995), and
(ii) it retains the positional information of SN\,1986J with respect 
to the phase-reference source, 3C\,66A, almost 1 degree apart.
Indeed, a Fourier inversion of the SN\,1986J data within DIFMAP, before applying
any self-calibration, showed that the a priori position of SN\,1986J
used in the correlator
($\alpha = 02\fh 22\fm 31\fs320$, $\delta = +42\fdg 19\farcm 57\farcs282$; J2000.0)
was off from the actual one by $\sim$17 mas in right
ascension (westwards), and by $\sim$21 mas in declination (southwards).  
Therefore, phase-referencing proved to be also useful in providing
further refinements in the precise coordinates of 
SN\,1986J.

\section[]{Results and Discussion}
\label{sec:results}

The only other available radio image of SN\,1986J is that obtained 
by Bartel et al. (1991; hereafter B91) of 29 September 1988, 
at a frequency of 8.4 GHz. 
Our global VLBI image of 21 February 1999 
was obtained at 5 GHz. 
The most remarkable thing in our image (panel (b) in Fig.~\ref{sn86j})
is the fact that the brightness distribution 
of SN\,1986J still shows the form of a shell, even though more
than 10 yr since the last observations have passed. 
However, the shell, definitely distorted and with strong asymmetries, 
shows changes in the brightness distribution between the two epochs.

\subsection{Asymmetric shell morphology and protrusions in SN\,1986J}

We distinguish at least four bright features in the rim of the shell
labeled A,B, C, and D in Fig.~\ref{sn86j}.
The absolute minimum ($\sim$0.26 mJy beam$^{-1}$) 
of the shell brightness distribution is indicated by 
the X in panel (b). 
The flux densities and distances of the bright features 
A,B,C, and D from X are shown in Table \ref{tab1}. 
At a flux density of 1.13 mJy beam$^{-1}$, the corresponding 
brightness temperature of feature A is $\sim1.8\, \times\, 10^{7}\,$\,K.   
Features B and C are local maxima with brightness temperatures of 
$\sim$1.0 and $\sim$1.2\, $\times\, 10^{7}$\,K, respectively.
Component D is an extended feature in the AC direction. 
Therefore, its flux density and radial distance from X are 
just indicative.

The image by B91 showed several `knots' along the rim of the shell  
of the supernova brightness distribution.  
The strongest knot was located northeast of the center, with a peak brightness 
temperature of $\sim3\, \times\, 10^8\, {\rm K}$.
B91 also showed the presence of protrusions in the brightness distribution of 
the supernova, at a distance of $\sim$\,1.9 mas from the centre. 
We note that the radial distances to features A, B, and C in 
Fig. 1 are comparable to the radial distances to the protrusions observed
by B91, and therefore one could be tempted to identify features A, B, and 
C with stationary clumps in the CSM of SN1986J. However, 
caution must be taken, since not only the position angles are different, but
also our synthesized beam is significantly larger than that of B91.
Our image also shows three spots outside the shell structure 
at contours above 3 times the rms noise off-source. 
These spots have distances and position angles with respect to X 
that are significantly different from those obtained by B91, and
therefore do not correpond to the same physical features shown in B91.
The peaks of these spots range from 0.23 to 0.25 mJy/beam, 
less than the minimum of emission within the shell. 
Therefore, they could be mere artifacts of the image reconstruction procedure,  
implying that the protrusions seen by B91 
have disappeared during the period ranging from September 1988 to 
February 1999, and pointing to a change in 
the density properties of the presupernova wind.  

Our image also shows that the radio shell morphology of SN\,1986J is asymmetric. 
The four bright features in Fig. \ref{sn86j} seem to delineate a highly distorted shell 
--whose minimum of emission is not at its centre-- and are 
indicative of a significant deformation of the shock front. 
If the point X in panel (b) corresponds to the centre of the explosion, 
we obtain the angular radial distances shown in Table \ref{tab1} 
for the knots in the shell of the supernova. 
It would appear then that the supernova has expanded more along the AC axis.
In addition, the angular radial distance XA is more 
than twice the distance XC, 
indicative of a strong asymmetric expansion. 
We note, however, that our identification 
of X with the centre of the supernova explosion 
is not unique. 
In this respect, future phase-reference VLBI observations will be most useful, 
since this technique will allow a proper registration of the maps, thus 
permitting the identification of the centre of the explosion. 

\begin{table}
\caption[]{Flux density and radial distance to X for the 
features shown in panel (b) of Fig.~\ref{sn86j}. 
The position of X is $\alpha = 02\fh 22\fm 31\fs 3214, \delta = 42\fdg 19\farcm 57\farcs 260$ 
}
\tabcolsep 7pt
\begin{tabular}{lcc} 
\hline  
          & Flux                & Radial distance  \\ 
          & mJy beam$^{-1}$     &  (mas)           \\
\hline 
A           & 1.13  & 2.14  \\
B           & 0.62  & 1.10  \\
C           & 0.77  & 0.99  \\ 
D           & 0.74  & 1.41  \\ 
\hline     
\end{tabular}
\label{tab1}
\end{table}

Nevertheless, most of the results presented and discussed below 
do not depend on the putative centre of the explosion. 
In particular, the brightness
distribution of SN\,1986J is clearly asymmetric, independent 
of the location of the centre of the explosion.
How could such an asymmetric structure have arisen?
Various models have been proposed to account for asymmetric
morphologies in supernovae.
Blondin, Lundqvist \& Chevalier \cite{blondin96} have suggested
that an axisymmetric density distribution in the wind
from a supernova progenitor leads to protrusions emerging along
the symmetry axis. 
These authors find that for a power-law supernova density profile,
the flow approaches a self-similar state in which the 
protrusion length is 2--4 times the radius of the shell, after 
$\sim10$\,yr. 
They cite the supernova remnant 41.95+575 in M\,82 as an example where 
such axisymmetric protrusions are likely to have emerged, but point out that
their model is not compatible with the protrusions seen by B91 for SN\,1986J.
A case where considerable asymmetry in 
its overall structure has been observed is SN1987A. 
Gaensler et al. \cite{gaensler97} showed
that the eastern and western regions of the SN1987A radio remnant 
were brighter than the northern and southern regions, and
interpreted this as evidence for the shocked wind being axisymmetric
in its shape and/or density distribution. 
From our image, it follows that 
an axisymmetric circumstellar interaction does not seem 
to apply for SN1986J. 
The protrusions outside the shell, on the other hand, 
could be formed by an asymmetric distribution in the 
wind similar to that invoked by Blondin et al. \cite{blondin96}, though not
restricted to be axisymmetric.
Other models that predict supernova aspherical morphologies include that of
Khokhlov et al. \cite{kho99}, which have modeled jet-induced explosions
of core collapse supernovae. The end result is a highly aspherical
supernova with two high-velocity jets of material moving in polar
directions, and a slower moving, highly distorted ejecta containing
most of the supernova material. 
A recent VLBI image of 41.95+975 by McDonald et al. 
\cite{mcdonald01} reveals structure resembling that of a
collimated outflow, as expected from the model by  
Khokhlov et al. \cite{kho99}.
Although in our case one could identify the brightest knots A and 
C with a collimated outflow, there is also strong emission in the
almost perpendicular direction BD. In addition, all four knots
seem well confined within the supernova shell. 
Therefore, the existence of two high-velocity, 
well collimated jets in SN\,1986J seems unlikely.
In the case of SN1987A, a directional anisotropy in the
supernova explosion might be indicated by the fact that
the eastern limb of SN1987A was brighter,  
and was also increasing in brightness more rapidly, 
than the western limb (Gaensler et al. 1997). 
Although such a directional anisotropy 
in SN1986J could explain the existence of feature A (which roughly
corresponds to the strongest knot of emission seen by B91), it has difficulties
in explaining the change in position angles for the other peaks, as well as the
appearance of features B and D.
We therefore favour an scenario where the strongly 
asymmetric brightness distribution of the 
SN\,1986J shell structure is due to the collision of the supernova ejecta 
with a clumpy (Chugai 1993, Chugai \& Belous 1999), 
or filamentary wind (Weiler et al. 1990).  
Our image gives support to this model, and 
shows that the clumpy, or filamentary  wind must probably be 
inhomogeneous to produce a highly distorted shell.

\subsection{Deceleration of the expanding shell}
\label{deceleration}

At a distance of 9.6 Mpc (Tully 1998), 1 mas corresponds to 
a linear size of $\sim1.4\, \times\, 10^{17}\,{\rm cm}\,\approx0.05\,{\rm pc}$.
Based on 8.4 GHz VLBI observations, 
B91  found an angular size  of $\sim$3.7 mas 
($\approx5.3\,\times\,10^{17}\,{\rm cm}\,\approx0.17\,{\rm pc}$)
for the shell of SN\,1986J, 5.74 yr 
after its explosion (assuming it took place on 1983.0). 
The corresponding mean linear velocity would then be
$\sim$14700\,km\,s$^{-1}$ for the first 5.74 yr.
However, this velocity applies only to the protrusions found by B91, not
to the shell. 
Indeed, the velocities reported in B91
were calculated for the protrusions, and assuming that these
originated in the centre at the time of the explosion.
These authors also pointed out that  the protrusions extended
from the centre to twice the radius of the shell, i.e., 
the protrusions were {\it outside} the shell. 
Therefore, a value of $\sim$1.85 mas 
($\approx2.7\,\times\,10^{17}\,{\rm cm}\,\approx0.09\,{\rm pc}$)
for the angular size of the shell of SN\,1986J
at epoch 1988.74 is indicated, and a mean linear velocity of the 
radio shell of $\sim$7400\,km\,s$^{-1}$ is more appropriate 
to characterize the first 5.74 yr of the expansion of SN\,1986J, 
as has been previously noticed by Chevalier \cite{che98} and Houck
et al. \cite{houck98}.  
Chevalier \cite{che98} also pointed out that such a velocity at 
roughly the time of the peak flux is evidence against the synchrotron
self-absorption mechanism acting in SN\,1986J.

The angular size of the supernova (at the 5-rms noise level) 
along the AC direction in Fig.\ref{sn86j} is 
$\theta \approx 4.7 \pm 0.3$ mas, 
equivalent to $\approx6.8\,\times\, 10^{17}\,$ cm $\approx\,0.22\,$ pc.
Combining this angular size measurement 
with that obtained by Bartel et al. (1991) for epoch 1988.74 
($\theta\,\approx\,1.85$\,mas), 
we then obtain a mean angular expansion velocity of the shell  
between 29 September 1988 (1988.74) and 21 February 1999 (1999.14) of
$\approx 0.14\,{\rm mas\, yr^{-1}}$, which corresponds to a linear
velocity of $\sim6300\,{\rm km\,s^{-1}}$.
If we assume that SN\,1986J freely expanded for the
first 5.74 yr of its life  ($r\,\propto\,t^m, m=1$) and
then started to decelerate, 
the expansion between the two epochs of VLBI observations
is characterized by $m = 0.90 \pm 0.06$. 
This is a mild deceleration, 
and contrasts with the case 
of several other supernovae, for which
a strong deceleration has been measured 
(SN1979C:  Marcaide et al. 2002;
SN1987A: Staveley-Smith et al. 1993, Gaensler et al. 1997; 
SN1993J: Marcaide et al. 1997, Bartel et al. 2000). 
Thus, the angular size that we obtain for the shell
seems consistent with the conclusion that 
the bulk of the shell structure expanded
for the first 5.74 yr at speeds smaller than $\sim 15000 \kms$
(see above and Chevalier 1998). 

For a standard presupernova wind velocity, $v_w$=10 \kms, 
the linear size of SN\,1986J at epoch $t$=16.14 yr implies that
we are sampling the progenitor wind about 11000 yr prior to
its explosion. 
Since the mass loss rate of SN\,1986J seems to have been
$\ga2\,\times\, 10^{-4}\,\msunyr$ (Weiler, Panagia \& Sramek 1990), 
the swept-up mass must have been ${\rm M_{sw}}\,\approx2.2\Ms$, and the thermal 
electron density $\sim8\times\,10^3\ccc$
(for a standard density profile of the progenitor wind, 
$\rho_{\rm cs} \propto r^{-2}$).
Since the expansion of the supernova has not decelerated significantly 
between the two VLBI observations, it follows that the swept-up 
mass by the shock front must be much less than the mass 
of the ejected hydrogen-rich envelope, ${\rm M_{env}}$,
as otherwise we should have observed a much stronger deceleration. 
In fact, momentum conservation implies that
${\rm M_{env}}\, \ga\, 12\, \Ms$, significantly larger than ${\rm M_{sw}}$. 
If the hydrogen-rich mass envelope was as high as 
12\,\Ms, this is a hint that the progenitor of SN\,1986J was probably a 
single, massive Red Super Giant (as previously suggested by Weiler et al. 1990), 
which lost mass rapidly, but managed to keep intact most of 
its hydrogen-rich envelope by the time of explosion.
This result contrasts with the cases of SN\,1993J and SN\,1979C, whose
hydrogen-rich envelopes had masses of 0.2--0.4\,\Ms 
(Woosley et al. 1994, Houck \& Fransson 1996) and $\la\,0.9\,\Ms$ 
(Marcaide et al. 2002), and whose progenitor stars were part of binary 
systems.

\subsection{Total energy and magnetic field of SN\,1986J}
\label{energy}

Since the radio emission is of synchrotron origin, 
we can estimate a minimum total energy (in magnetic fields,
electrons, and heavy particles) and a minimum magnetic field for SN\,1986J. 
If we assume equipartition (magnetic field energy is approximately equal 
to the total particle energy), then  
the minimum total energy is (Pacholczyk 1970)
$E_{\rm min}^{\rm (tot)} = c_{13}\, (1 + k)^{4/7}\, \phi^{3/7}\, R^{9/7}\, L^{4/7},$ 
where $L$ is the radio luminosity of the source,
$R$ is a characteristic size,
$c_{13}$ is a slowly-dependent function of the spectral index, $\alpha$, 
$\phi$ is the fraction of the supernova's volume occupied  by the magnetic field and 
by the relativistic particles (filling factor), and
$k$ is the ratio of the (total) heavy particle energy to the electron energy.
This ratio depends on the mechanism that generates the relativistic electrons, ranging
from $k \approx 1$ up to $k = m_p/m_ e \approx 2000$, where $m_p$ and $m_e$  
are the proton and electron mass, respectively. 

Based on snapshot VLA observations of SN\,1986J carried out on 13 June 1999 (Weiler, private
communication), we determined an spectral index of
$\alpha =-0.69 \pm 0.06$ ($S_\nu \propto \nu^{\alpha}$) between 1.6 and 8.5 GHz.
This value is very close to $\alpha = -0.67$ obtained by B91, 
and is a typical value for supernovae that are in the optically thin radio regime.
With $\alpha =-0.69$ and $S_{4.9 \rm GHz} = 7.2$ mJy from our observations, we obtain 
a radio luminosity $L_\nu \approx 1.38 \times 10^{37} (D/9.6 {\rm Mpc})$ erg\,s$^{-1}$ 
for the frequency range  $10^7$--$10^{10}$ Hz. 
The value of the function $c_{13}$ is approximately $2.39 \times 10^4$
(for $\alpha =-0.69$, and $\nu_1$ and $\nu_2$ equal to $10^7$ and $10^{10}$ Hz,
respectively). 
The filling factor, $\phi$, is by definition $\leq$ 1. 
In the case of 
supernova SN\,1993J, Marcaide et al. (1995a, 1995b, 1997) 
showed that the radio emitting shell of SN\,1993J had a width, $\Delta\,R \approx 0.3\, R_{\rm out}$, 
where $R_{\rm out}$ is the outer shell radius, and therefore $\phi \approx 0.66$.
As the characteristic size for SN\,1986J, we will take half the largest 
diameter of the shell, $\theta = 2.35\, {\rm mas}$, which corresponds 
to a linear size of $R = D\, \cdot\, \theta\, \approx3.4\, \times 10^{17}$cm.
With these values, the minimum total energy is then 

\[
 E_{\rm min}^{\rm (tot)} \approx 1.13 \times 10^{48} \,(1 + k)^{4/7}\, 
   \phi_{0.66}^{3/7}\,
   \theta_{2.35}^{9/7}\,   
   D_{9.6}^{14/7}\,  {\rm erg}
\]

\noindent
where $ \phi_{0.66} = (\phi/0.66)$
$ \theta_{2.35} = (\theta/2.35\,{\rm mas})$ , and
$ D_{9.6} = (D/9.6 {\rm Mpc})$.
The value of the magnetic field that yields $E^{\rm (min)}_{\rm tot}$ is
then equal to

\[
 B_{\rm min} \approx 10.6 \, (1 + k)^{2/7}\, 
   \phi_{0.66}^{-2/7}\, 
   \theta_{2.35}^{-6/7}\,   
   D_{9.6}^{-2/7}  {\rm mG}
\]

\noindent
Since $1 \leq k \leq 2000$,  
$E_{\rm min}^{\rm (tot)}$ can have values between $\sim 2 \times 10^{48}$ 
and $\sim 9 \times 10^{49} {\rm erg}$, while the corresponding values of the
magnetic field can lie between $\sim 13$  and
$\sim 93\, {\rm mG}$ (for $\phi = 0.66$ and $D = 9.6 {\rm Mpc}$).  
These values of the magnetic field for SN\,1986J are in agreement
with, e.g., those obtained for SN\,1993J at similar radii.
For example, P\'erez-Torres, Alberdi, and Marcaide \cite{ptorres01}
showed that $B\approx\,0.58\, (r/3\,\times\,10^{16}\,{\rm cm})^{-1}$\,G, 
which results in a value of $\sim\,50$\,mG for a radius of 
$3.4\,\times\,10^{17}$\,cm.
Since it is very unlikely that the magnetic field energy density in the
wind is larger than its kinetic energy density, i.e.,
$B^2/8\pi \leq \rho v^2_{\rm w}/2$, we can 
then obtain an upper limit for the magnetic field in the circumstellar
wind of SN\,1986J:

\[
 B_{\rm cs} \la (\mdot\,v_w)^{1/2}\,r^{-1}\,\approx\,
                        0.78\,(\mdot_{-4}\,v_{10})^{1/2}\, r_{17}^{-1}\,{\rm mG} 
\]

\noindent
where $\mdot_{-4} = \mdot/10^{-4}\,\msunyr$, $v_{10}\,=\,v_w/10\,\kms$, 
$r_{17} = r/10^{17}\,{\rm cm}$, and  
we have assumed a standard wind density profile.
For SN\,1986J, $\mdot = 2\,\times\,10^{-4}\,\msunyr$ and $r\,=\,3.4\,\times\,10^{17}$\,cm, 
and we obtain $B_{\rm cs}\, \la \,0.32$\,mG. 
These arguments suggest that the magnetic field 
in the shell of SN\,1986J is in the range 13--93 mG, or about 40 to 300 times 
the magnetic field in the circumstellar wind. 
Since a strong shock yields a fourfold increase in the 
particle density, the post-shock magnetic field is 
also fourfold increased.
Hence, compression alone of the circumstellar wind magnetic field cannot
account for the large magnetic fields existing in SN\,1986J, and 
other field amplification mechanisms are needed to be invoked, e.g., 
turbulent amplification (Chevalier 1982; Chevalier \& Blondin 1995). 
The same conclusion was reached for the case of SN\,1993J 
(Fransson \& Bj\"ornsson 1998, P\'erez-Torres, Alberdi \& Marcaide 2001), 
where magnetic field amplification factors $\sim100\,$ were found to be necessary.

\section[]{summary}
\label{sec:conclusions}

We used 5 GHz global VLBI observations of SN\,1986J taken on 21 February 1999
to obtain a high-resolution radio image of the
supernova about 16 yr after its explosion (assumed to have occurred
on 1983.0). Our main results can be summarized as follows:\\

(i) The radio structure of SN\,1986J is shell-like, 
even after more than 16 yr since its explosion. 
However, the shell is highly distorted, indicative of a strong 
deformation of the shock front. 

(ii) 
Assuming that the shell seen in our 1999.14 image 
corresponds to the shell in the 1988.74 image, 
the expansion of SN\,1986J 
has decelerated from $\sim$7,400 \kms\ down 
to $\sim$6,300 \kms. 
This mild deceleration can be characterized by a power-law ($r \propto t^m$) 
with a deceleration parameter $m=0.90\,\pm\,0.06$.
Since the swept-up mass must be $\approx2.2\,\Ms$
(for $\mdot =2\,\times\, 10^{-4}\,\msunyr$ and $v_w=10\, \kms$), 
this mild deceleration seems to indicate 
a mass of the hydrogen-rich envelope ejected at explosion
of $\ga\, 12\,$\Ms.  In this case, 
the supernova progenitor kept almost intact its hydrogen-rich envelope, 
in spite of its strong mass-loss wind rate, and is 
a strong hint that SN\,1986J went off in a single, 
massive star scenario.

(iii) We show that the brightness distribution in the shell is strongly asymmetric, 
and suggest that the most likely scenario for its origin is  
the collision of the supernova envelope with a clumpy, or filamentary, 
anisotropic wind. 
However, we cannot exclude that other mechanisms, 
such as a directional anisotropy in the supernova
explosion, may also play a role in shaping the radio emitting structure of SN1986J.
We detect three emission spots at the 3-rms noise level outside the shell.
Though these spots could be protrusions alike those detected by Bartel et al. (1991), 
we suggest that they are mere artifacts of the image reconstruction
procedure. 

(iv) The radio flux of SN\,1986J is $\sim7$ mJy, which results in a  luminosity of
about $1.4\,\times\, 10^{37}\, {\rm erg\,s^{-1}}$.  By assuming equipartition 
between fields and particles, we estimate a minimum energy for the supernova shell 
in the range $E_{\rm min}^{\rm (tot)} \approx$ 0.2--9 $\,\times\, 10^{49}\,$ erg, 
which corresponds to a minimum magnetic field of $B_{\rm min} \approx$ 13--93\,mG.
We find that the average magnetic field in the circumstellar wind of SN\,1986J
is $B_{\rm cs}\,\la0.32$\,mG. Therefore, powerful amplification
mechanisms (e.g. turbulence) should be acting in SN\,1986J, since compression 
alone of the circumstellar field cannot account for the magnetic fields 
existing in the supernova shell.

The high-resolution radio image of SN\,1986J presented here shows that 
VLBI is a very powerful technique for investigating the interaction
of supernovae, because it allows to directly image the emission from the
interaction region (supernova ejecta and shocked wind).  
Supernova SN\,1986J belongs to the rare group of
radio supernovae that permit their follow-up with VLBI.   
Further high-resolution, sensitive VLBI observations of SN\,1986J 
are necessary to provide us with a more detailed view of its evolution and 
interaction with the circumstellar medium. 
The use of phase-referencing will be mandatory, since
this technique will not only ensure the detection of the supernova, 
but will also allow a proper registration of the images for all epochs.
This registration will likely permit the unambiguous identification
of the centre of the explosion.
Moreover, such observations would also allow to shed further light on the apparent 
asymmetric presupernova wind of SN1986J, 
obtain a better determination of the deceleration parameter, and
definitively confirm, or reject, the protrusions outside
the supernova shell structure we have marginally detected here.

\section*{Acknowledgments}
We thank the National Radio Astronomy Observatory (NRAO) for providing
us with the archival data used in this paper.
We thank Roger Chevalier, Nikolai Chugai, and Alan Roy 
for comments on the manuscript.
We are grateful to an anonymous referee for many substantive and useful 
comments, which significantly improved our manuscript.
This research has been supported by a Marie Curie Fellowship
of the European Community (contract IHP-MCFI-99-1),  
and by the Spanish DGICYT grants AYA2001-2147-C02-01/02. 
KWW thanks the Office of Naval Research (ONR) for the 6.1
funding supporting this research.
The NRAO is a facility of the National Science
Foundation operated under cooperative agreement by Associated Universities, 
Incorporated.
The European VLBI Network is a joint facility of European and Chinese
radio astronomy institutes funded by their national research councils.

\end{document}